\documentclass[a4paper,10pt,twoside]{cpc-hepnp}

\usepackage{multicol}
\usepackage{graphicx}
\usepackage{booktabs}
\usepackage{amssymb,bm,mathrsfs,bbm,amscd}
\usepackage[tbtags]{amsmath}
\usepackage{lastpage}
\usepackage{subfigure}

\begin{document}

\fancyhead[co]{\footnotesize Conception design of helium ion FFAG accelerator with induction accelerating cavity}


\title{Conception design of helium ion FFAG accelerator with induction accelerating cavity\thanks{Supported by National Natural Science Foundation of China (11045003.) }}

\author{%
      Luo Huan-li(ÂÞ»ÀÀö)$^{1;1)}$\email{huanli@mail.ustc.edu.cn}%
\quad Xu Yu-cun(ÐìÓñ´æ)$^{1,2}$
\quad Wang Xiang-qi(ÍõÏàôë)$^{1;2)}$\email{wangxaqi@ustc.edu.cn(Corresponding author)}%
\quad Xu Hong-Liang(ÐìºêÁÁ)$^{1}$
}
\maketitle

\address{%
$^1$ National Synchrotron Radiation Lab., University of Science and Technology of China, Hefei 230029, China\\
$^2$ China Electronics Technology Group Corporation No.38 Research Institute, Hefei 230088, China\\
}

\begin{abstract}
In the recent decades of particle accelerator R\&D area, fixed field alternating gradient (FFAG) accelerator has become a highlight for some advantages of its higher beam intensity and lower cost, although there are still some technical challenges. In this paper, FFAG accelerator is adopted to accelerate helium ion beam on the one hand for the study of helium embrittlement on fusion reactor envelope material and on the other hand for promoting the conception research and design of FFAG accelerator and exploring the possibility of developing high power FFAG accelerators. The conventional period focusing unit of helium ion FFAG accelerator and three-dimensional model of the large aperture combinatorial magnet by OPERA-TOSCA are given. For low energy and low revolution frequency, induction acceleration is proposed to replace conventional radio frequency(RF) acceleration for helium ion FFAG accelerator, which avoids the potential breakdown of acceleration field caused by wake field and improves the acceleration repetition frequency to gain higher beam intensity. The main parameters and three-dimensional model of induction cavity are given. Two special constraint waveforms are proposed to refrain from particle accelerating time slip($\Delta$T) caused by accelerating voltage drop of flat top and energy deviation. The particle longitudinal motion in two waveforms is simulated.
\end{abstract}



\begin{multicols}{2}

\section{Introduction}
With further development of some related technologies and three-dimensional magnetic field computing softwares, fixed field alternating gradient (FFAG) accelerator has again become a highlight in particle accelerator R\&D area fifty years later, depending on higher average beam intensity, higher repetition frequency, small volume and low cost. FFAG accelerator is a class of circular particle accelerator with fixed magnetic field and alternating gradient focusing, combining the advantages of separated strong-focusing accelerator and weak-focusing cyclotron. It has the characters of large radial acceptance and momentum acceptance due to outward spiral orbits along the radial direction and high momentum compaction factor, and also the repetition frequency is limited only by the accelerating cavity. The FFAG accelerator has such wide applications as breeding nuclear materials and transmutating nuclear wastes in accelerator driven system(ADS)\cite{lab1,lab2,lab3}, muon acceleration and neutrino factory in secondary beam facilities\cite{lab4,lab5,lab6} and cancer therapy and 3D spot scanning irradiation technique in medical application\cite{lab7,lab8,lab9}.

The structural material with some characteristics of high thermostability, high radioresistance and anti helium embrittlement is a bottleneck of the whole nuclear power project to realize safe and reliable operation\cite{lab10,lab11}. The mechanism study of helium embrittlement on fusion reactor envelope material and the method to delay the influence caused by helium embrittlement on construction material are very necessary and significant. Therefore, FFAG accelerator is used to provide helium ion beam for the study of helium embrittlement and also for promoting the deeper understanding of FFAG accelerator conception and exploring the possibility of developing high power FFAG accelerator.

In this paper, induction acceleration\cite{lab12,lab13} replaces radio frequency (RF) acceleration widely adopted by some present running FFAG accelerators\cite{lab1,lab4}. Compared with RF cavity, the induction cavity avoids the acceleration field breakdown caused by the wake field and improves instantaneous beam intensity as high as more than 10 kA\cite{lab14}. In addition, compared with the circular induction accelerator (CIA), the space charge effect is weakened, because the beam size can be stretched along the radial and longitudinal directions for beam multi-orbits in FFAG accelerator.

\section{Linear beam optics design of helium ion FFAG accelerator}
Considering ion energy in the helium embrittlement and the relation between the energy and the ion penetration depth in the fusion reactor envelope material, helium ion with two positive charges during 2$-$36 MeV is provided by the FFAG accelerator. Helium ion FFAG has the parameters of orbit displacement $0.5$ m and the repetition frequency $500$ Hz.

The triplet scheme of the radial scaling FFAG accelerator is selected. There are stable magnetic field configuration, well beam optics properties and small edge field effect in the radial scaling FFAG. Furthermore, the long straight section is very favorable for particle injection and extraction, accelerating cavity and beam diagnosis.

\begin{center}
\includegraphics[angle=0,width=8.8cm]{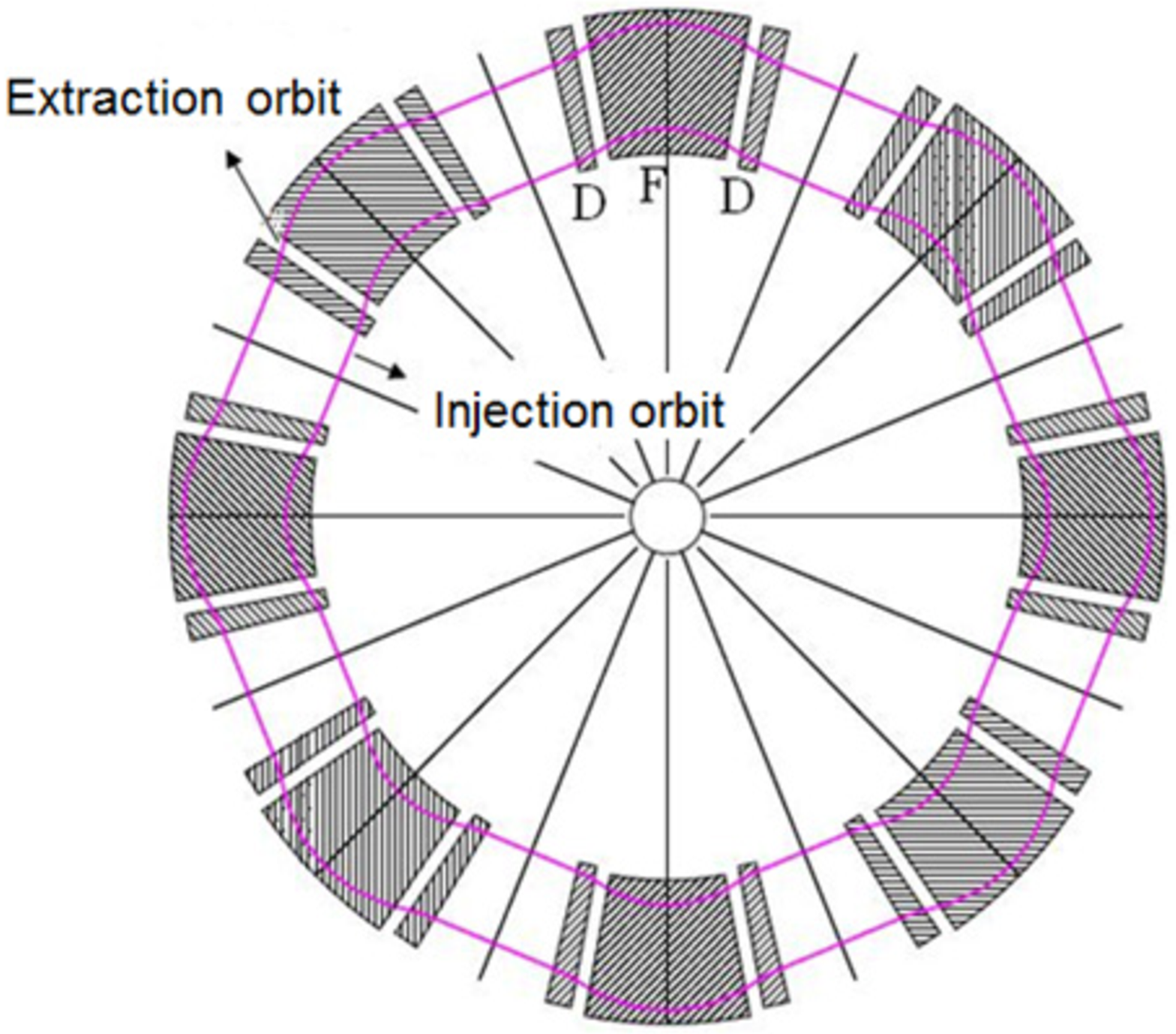}
\figcaption{\label{fig1} A layout of super-period structure for helium ion FFAG accelerator. The focusing F iron and defocusing D iron respectively make particle forward deflection and reverse deflection. The overall dimension is about 2.7 m$\times$2.7 m. The beam diagnostic and accelerating cavity are not displayed.}
\end{center}

The eight super-periods DFD and H type bending magnet with field index $4.44$ are considered for helium ion FFAG accelerator and the preliminary super-period structure plane is shown in Fig.~\ref{fig1} with the main parameters listed in Table~\ref{tab1}. In the figure, the ideal closed orbits respectively corresponding to the injection and extraction are found showing the well orbit geometric similarity. In order to weaken the field interference, an $8$ cm straight drift section separates the focusing F iron from the defocusing D iron. The four induction cavities with accelerating voltage $5$ kV are installed in the preliminary design and the one accelerating per two laps can reduce the output repetition frequency requirement of modulator. Multi-turn injection also improves the average beam intensity. The betatron function of linear beam optics for injection is given in Fig.~\ref{fig2}. Particles don't run across the main resonance lines from injection to extraction, which are shown in Fig.~\ref{fig3}.

\begin{center}
\includegraphics[angle=0,width=8.0cm]{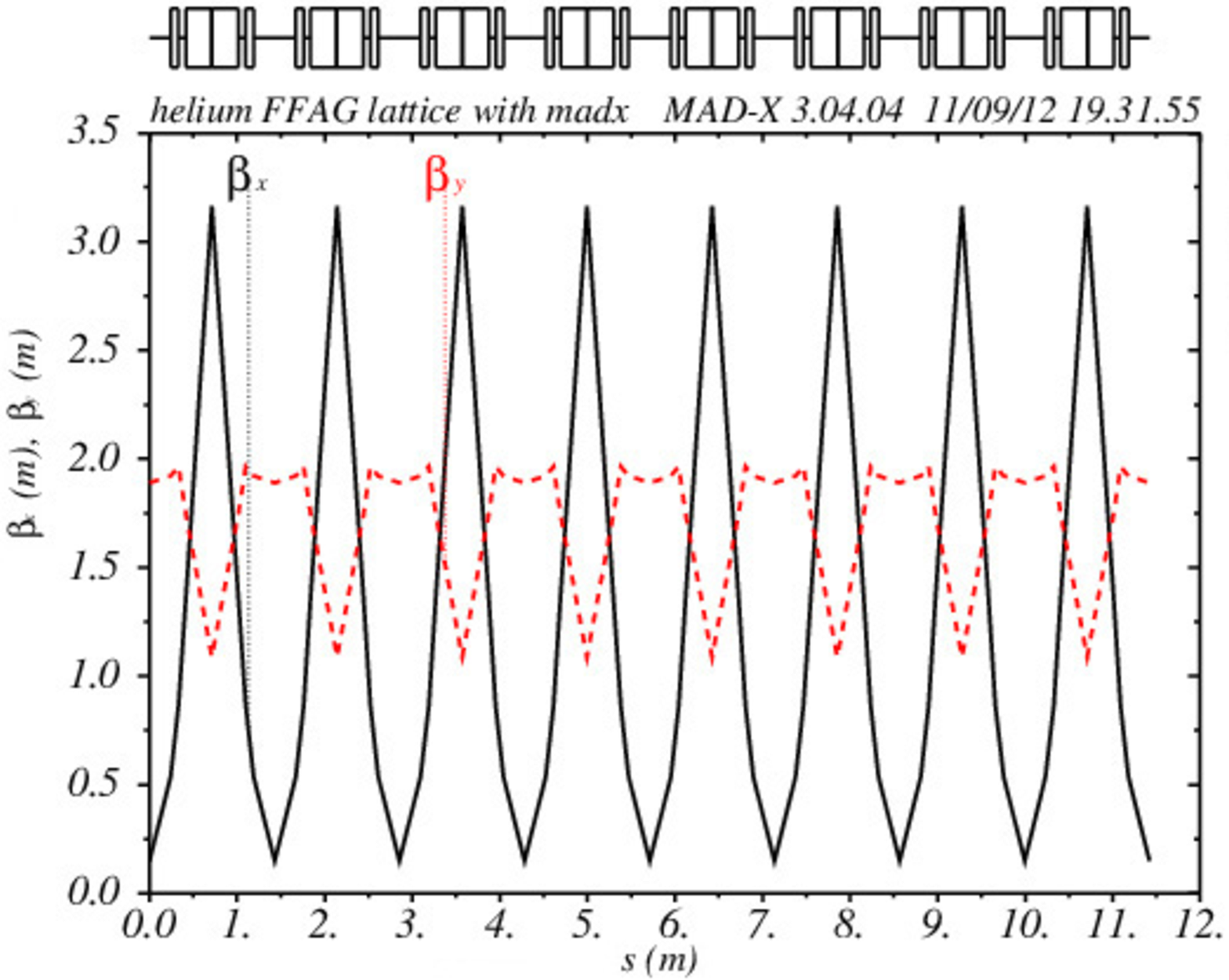}
\figcaption{\label{fig2} Betatron function of helium ion FFAG accelerator for injection.}
\end{center}
\vspace{-5mm}

\begin{center}
\includegraphics[angle=0,width=8.0cm]{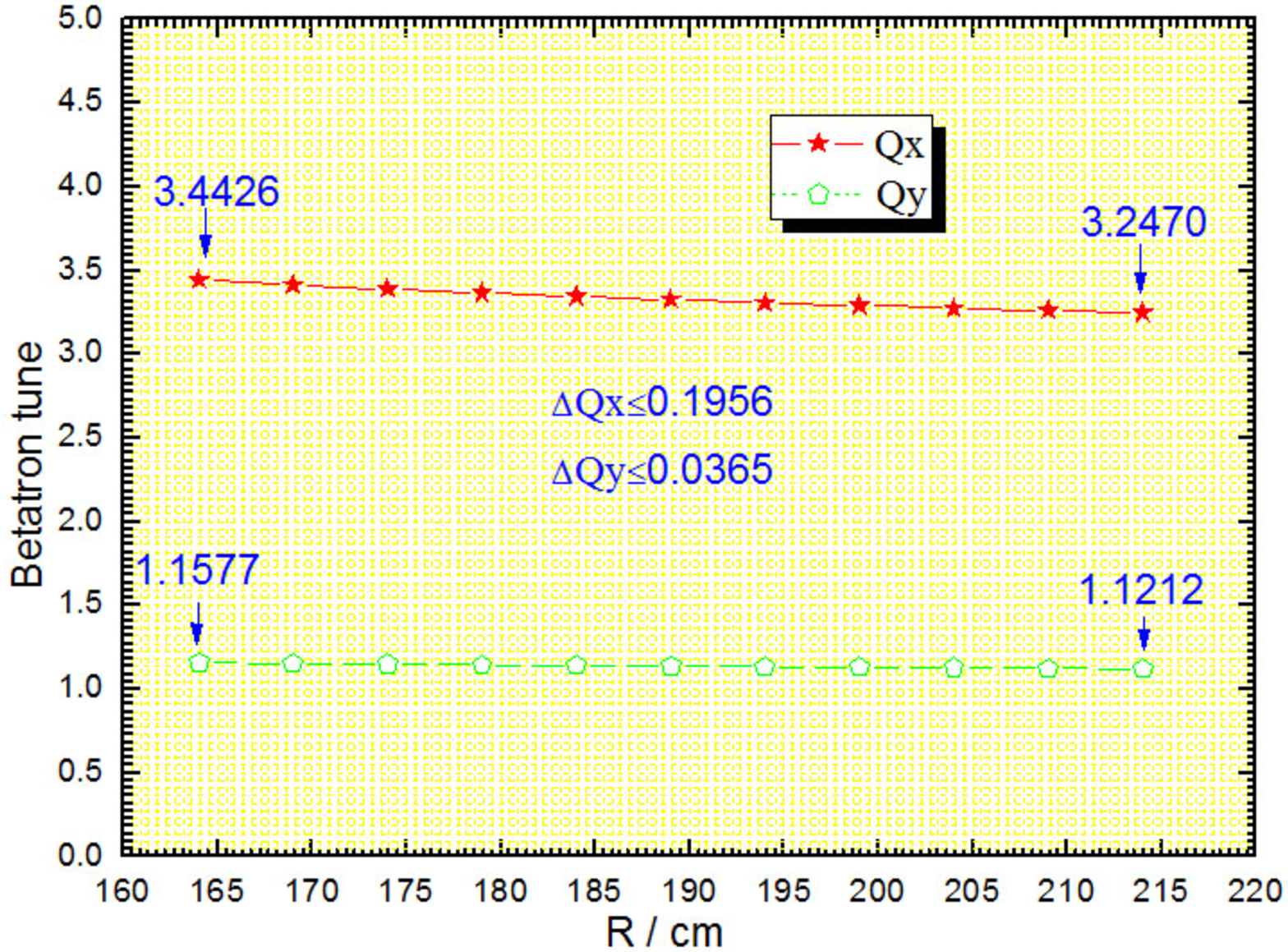}
\figcaption{\label{fig3} Tune drift from injection to extraction for helium ion FFAG accelerator.}
\end{center}
\vspace{-5mm}

\end{multicols}
\begin{center}
\tabcaption{ \label{tab1}  Some design targets and main parameters of helium ion FFAG accelerator.}
\footnotesize
\begin{tabular*}{140mm}{c@{\extracolsep{\fill}}ccc}
\toprule Parameter & Design goal & Parameter & Design goal \\
\hline
Particle & He$^{+2}$ & Energy$/$MeV & 2$-$36 \\
Magnet type & DFD & Super-periods & 8 \\
Field Index & 4.44 & Ring circumference$/$m & 11.42$-$14.51 \\
Orbit excursion$/$m & 0.50 & Revolution period$/$ns & 1163.60$-$350.90 \\
Curvature for F$/$m & 0.55$-$0.72 &  Curvature for D$/$m & -0.59$-$-0.77\\
F$/$2 Opening angle$/$deg. & 10.44 & D Opening angle$/$deg. & 3.20 \\
Long straight$/$m & 0.48$-$0.62 & Aperture$/$mm & 680(H)$\times$70(V) \\
Bending angle$/$deg.  & F$/$2$:$ 31.3 & Betatron tunes & In$:$ 3.44$/$1.16 \\
  & D$:$ 8.8 & & Fin$:$ 3.25$/$1.12 \\
Injection & Multi-Turn Injection & Cavitiy & 4 \\
Repetition frequency$/$Hz & 500 & energy gain per lap$/$keV & 40.0 \\
Average current extracted$/$mA & 10 & cavity voltage$/$kV & 5.0 \\
\bottomrule
\end{tabular*}
\end{center}
\begin{multicols}{2}

\section{The study of longitudinal motion with induction acceleration}
In conventional alternating gradient accelerator, improving particle energy depends on synchronously modulating magnetic field of main magnets and accelerating cavity frequency. However, for the fixed magnetic field in FFAG accelerator, the accelerating repetition frequency is limited only by the pulse accelerating waveform repetitive rate. The radial excursion of particle orbit in FFAG accelerator also brings about accelerating cavity with large enough horizontal aperture.

At present, the pulsed power technology of solid-state modulator is sufficient to meet the particle revolution period ranging from $1163.63$ ns to $350.9$ ns, respectively corresponding to the injection energy $2$ MeV to the highest extraction energy $36$ MeV. So, induction acceleration is proposed for helium ion FFAG accelerator, instead of radio frequency (RF) acceleration used by some present running FFAG accelerators. Induction acceleration adopted to helium ion FFAG accelerator has many advantages. There is no accelerating field breakdown caused by the wake field in induction cavity and then the beam instantaneous intensity is more higher. It's possible that a large enough horizontal aperture can further relax the limitation of space charge effect and also the horizontal coupled impedance of cavity can be ignored. Furthermore, the low energy helium ion FFAG accelerator doesn't need a too high cavity voltage and then there is no need to consider the high-voltage insulation and complicated accelerating gap, due to multiorbit acceleration, which is different from a linear induction accelerator(LIA). Moreover, the shielding body among each super-period structure weakens the influence which magnetic edge field has on shunt impedance of induction cavity. However, the cavity design, accelerating waveform and accelerating field distribution in induction cavity deserve further consideration.

\subsection{The design of induction cavity}
The racetrack structure of induction cavity in a beam transverse section depends on the vacuum chamber section (680 mm$\times$60 mm) of helium ion FFAG accelerator. The four induction cavities with cavity voltage $5$ kV are arranged and the helium ion is accelerated per two laps. The accelerating voltage $U_{{\rm c}}$ of induction cavity can be expressed as
\begin{eqnarray}
  U_{\rm c}  = -\int_S{({\rm d}B/{\rm d}t){\rm d}S},
\end{eqnarray}
where $B$ is the magnetic induction in the magnet, $S$ is a cross-sectional area of cavity. The racetrack induction cavity is regarded as a single turn coil and its inductance deciding accelerating voltage drop of flat top can be obtained by
\begin{eqnarray}
  L  = \frac{\mu\mu_{0}h}{2\pi}\ln\frac{2\pi r_{o}+2l}{2\pi r_{i}+2l}  ,
\end{eqnarray}
where $r_{i}$, $r_{o}$ correspond to the inner and outer radii of the semi-circle at both ends of the cavity, $l$ is the length in the linear part of cavity, $h$ is the thickness of the cavity along the beam longitudinal direction.

The preliminary design value of some main parameters for the induction cavity is listed in Table~\ref{tab2}. One assumes that the modulator output waveform is ideal with the fixed pulse width of $200$ ns. The average output beam intensity is $10$ mA and the particle extraction period is $2$ ms corresponding to the instantaneous beam intensity of $100$ A. The accelerating(output pulse) repetition frequency of modulator is within about 430 kHz and 1425 kHz.

\end{multicols}
\begin{center}
\tabcaption{ \label{tab2}   The main parameters of induction cavity in helium ion FFAG accelerator.}
\footnotesize
\begin{tabular*}{140mm}{c@{\extracolsep{\fill}}ccc}
\toprule Parameter & Design goal & Parameter & Design goal \\
\hline
Core material & FT-3L & Equivalent inductance$/$mH & 249.5 \\
Core straight length$/$mm & 680 & Equivalent capacitance$/$pF & 417.3 \\
Core inner diameter$/$mm & 90 & Core equivalent loss resistance$/$W & 280.13 \\
Core outside diameters$/$mm & 590 & Pulse width$/$ns & 200 \\
Core thickness$/$mm & 30 & Rise time$/$ns & 22.5 \\
Core number & 8 & Drop of flat top & 1.0$\%$ \\
Inductance in vacuum$/$nH & 31.65 & Duty cycle &  about 18$\%$ \\
Cavity voltage$/$kV & 5 & Beam instantaneous current$/$A & 100 \\
Accelerating gap$/$mm & 20 & Coaxial cable impedance$/$$\Omega$ & 25 \\
Repetition frequency$/$kHz & 430 & Matching load impedance$/$$\Omega$  & 60.86 \\
Power consumed by beam$/$kW & 90 & Core loss power$/$kW & 16.06 \\
Matching load loss power$/$kW & 73.94 & Total power$/$kW & 180 \\
\bottomrule
\end{tabular*}
\end{center}
\begin{multicols}{2}

The OPERA$-$3D model of accelerating cavity is shown in Fig.~\ref{fig4}. Here the integral value of electric field along the beam longitudinal direction is mainly investigated. The integral value of electric field (accelerating voltage) along $z$ axis (the beam horizontal direction) on $y = 0$ plane and $y = 2$ plane respectively is shown in Fig.~\ref{fig5}, where the integrating range is from $x = -20$ cm to $x = 20$ cm. As shown in Fig.~\ref{fig5}, the variation of accelerating voltage $U_{{\rm c}}$ along $z$ axis is about within 1$\%$ and the distributions of accelerating voltage on two considered planes ($y = 0$ plane and $y = 2$ plane) are basically the same, that is around $19$ V. The accelerating voltages at both ends of $z$ axis decline markedly, respectively corresponding to the injection orbit and extraction orbit, which can be compensated by increasing the radial span of induction cavity.

\begin{center}
\includegraphics[angle=0,width=7.0cm]{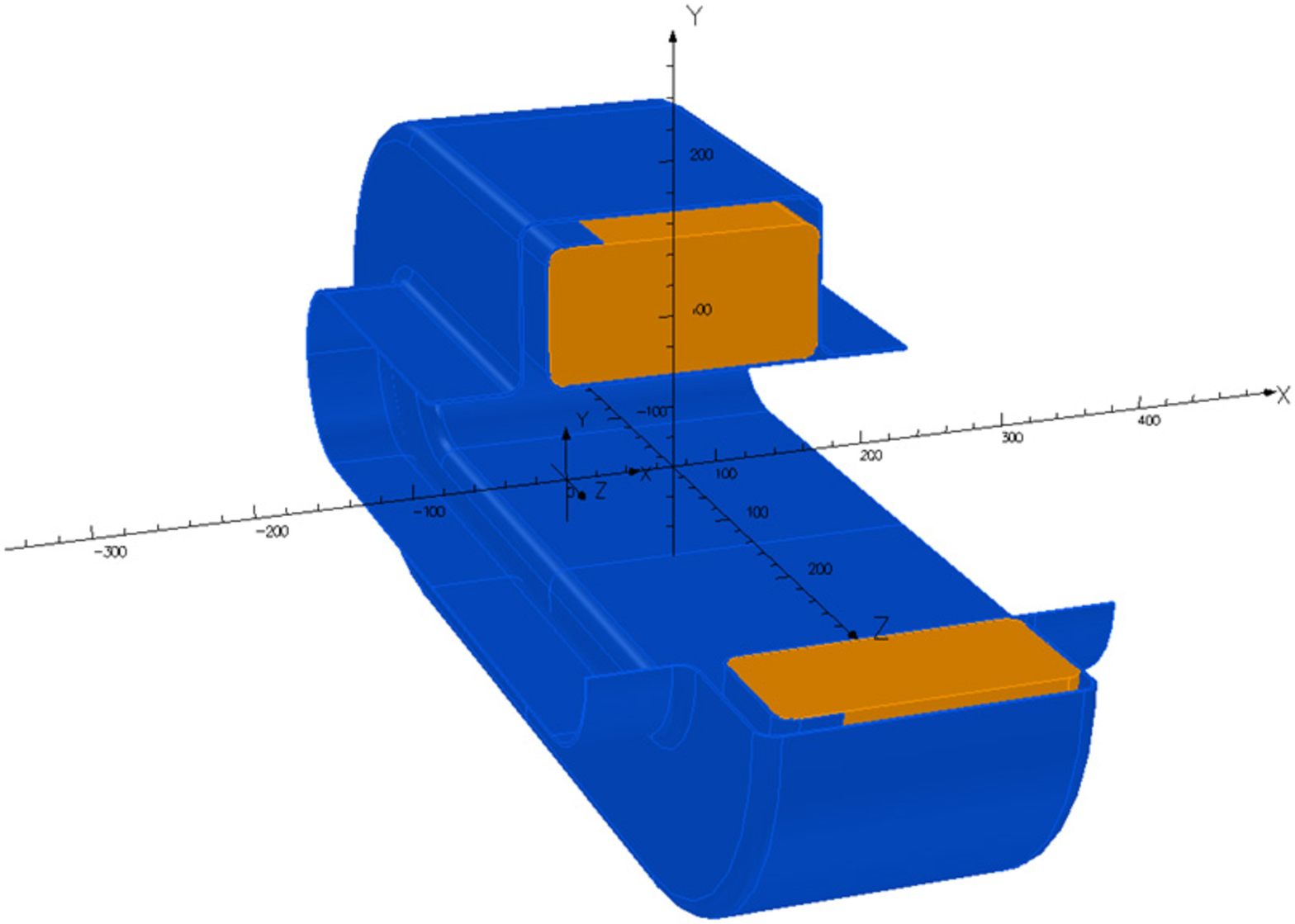}
\figcaption{\label{fig4} The OPERA-3D model of helium ion FFAG induction-cavity. The radial range of 3D model is $50$ cm, the vertical height is $10$ cm and the accelerating gap is $2$ cm. The $x$ axis in the figure is defined as the beam longitudinal direction, $y$ axis the beam vertical direction and $z$ axis the beam horizontal direction.}
\end{center}
\vspace{-4mm}

Furthermore, to adopt the circulating cooling dielectric oil system is to solve the problem of cavity core power loss. The matching load impedance connected in parallel with the induction cavity is about 60.86 $\Omega$. Since single solid-state modulator is difficult to achieve average beam output power above $100$ kW, many solid-state modulators can operate in parallel for helium ion FFAG accelerator with the higher average output beam intensity and output power.
\begin{center}
\includegraphics[angle=0,width=8.2cm]{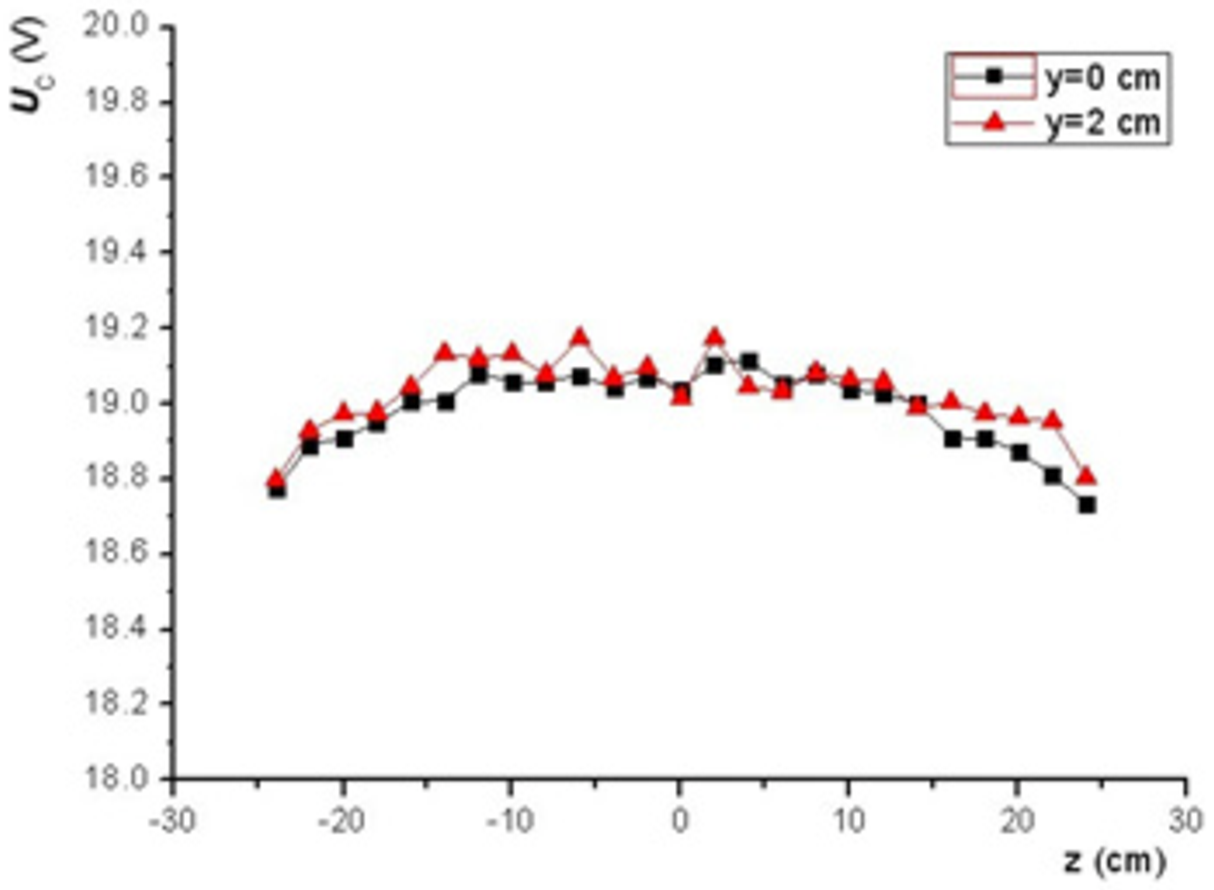}
\figcaption{\label{fig5} The acceleration voltage distribution on the middle plane of induction cavity.}
\end{center}
\vspace{-4mm}

\subsection{The simulation of longitudinal motion with energy compensation}
The resonance condition of modulation acceleration must be met for driving the FFAG induction acceleration, that is $T_{M}= mT_{N}$, where $T_{N}$(ns) is the revolution period of helium ion, $T_{M}$(ns) is the output pulse period of modulator, and $m$ represents one accelerating per $m$ laps, $m=2$ for helium ion FFAG accelerator. For certain beam optics, the revolution period of helium ion can be expressed as
\begin{eqnarray}
  T_{N} = \frac{A_{1}(\sqrt{(E_{{\rm in}}+n\Delta E)(E_{{\rm in}}+n\Delta E +2\varepsilon_{0})})^{\alpha}+A_{2}}{\sqrt{1-[{\varepsilon_{0}}/(\varepsilon_{0}+E_{{\rm in}}+n\Delta E)]^{2}}}
\end{eqnarray}
where $A_{1}$, $A_{2}$ are two parameters relating to beam optics, $A_{1}=13.98$, $A_{2}=4.27$ for Table~\ref{tab1}. $\alpha$ is a constant, $\alpha$ =0.18384. $E_{{\rm in}}$ is the injection energy $2$ MeV, and $\Delta$$E$ is the energy gain per each lap. The total accelerating number (pulse number) from particle injection to extraction is $850$.

\begin{center}
\includegraphics[angle=0,width=6.0cm]{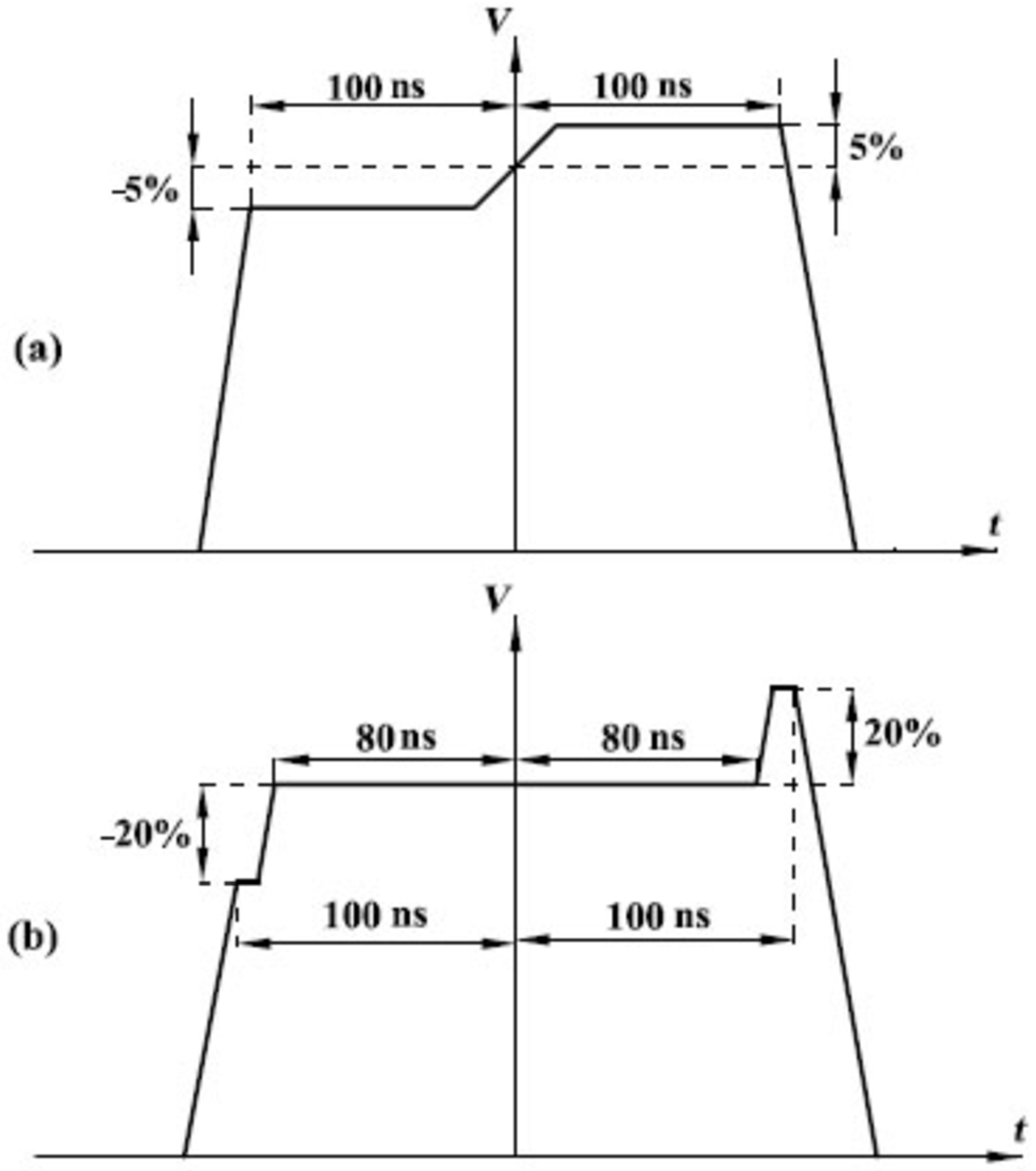}
\figcaption{\label{fig6} The schematic diagram of two acceleration voltage constraint waveforms with energy compensation: (a) the double platform waveform; (b) the head-tail restraint waveform.}
\end{center}
\vspace{-4mm}

The above discussion directs to some particles on the ideal condition. Some other particles on the nonideal condition, particularly the accelerating phase lag, are actually likely to break away from the accelerating phase area, owing to the induced voltage drop of flat top. For example, the time slip($\Delta$T) of a particle with 1$\%$ below the ideal accelerating energy is simulated in the study. When this particles energy is 36 MeV, it lags about 1.6 s behind the ideal particle, which ran out of the accelerating phase area and was lost long ago. So, some restraints need to be used to guarantee small enough accelerating time slip($\Delta$T) and then the helium ion does not escape from the accelerating phase area.

In order to solve the problem mentioned above, two constraint waveforms with energy compensation are proposed. One considers that the output pulse waveform of modulator has this shape shown in Fig.~\ref{fig6}. The waveform in Fig.~\ref{fig6} (a) is a double platform waveform found by the first half voltage reduced by 5$\%$ and the second half voltage raised by 5$\%$ on the basis of a flat-topped wave. The double platform waveform may be actually obtained by halving a certain pulse width of inductive adder solid-state modulator. The waveform in Fig.~\ref{fig6} (b) is a head-tail restraint waveform also improved on the basis of a flat-topped wave.

The particle longitudinal motion in two waveforms is simulated by Matlab coding. Some particles with different initial times and different initial energies are studied without the consideration of space charge effect. The evolutionary procedure of time slip($\Delta$T) and energy deviation during particle acceleration is discussed, as well as the longitudinal acceptance. The results are shown in Fig.~\ref{fig7}.

\end{multicols}
\begin{center}
\includegraphics[angle=0,width=14.0cm]{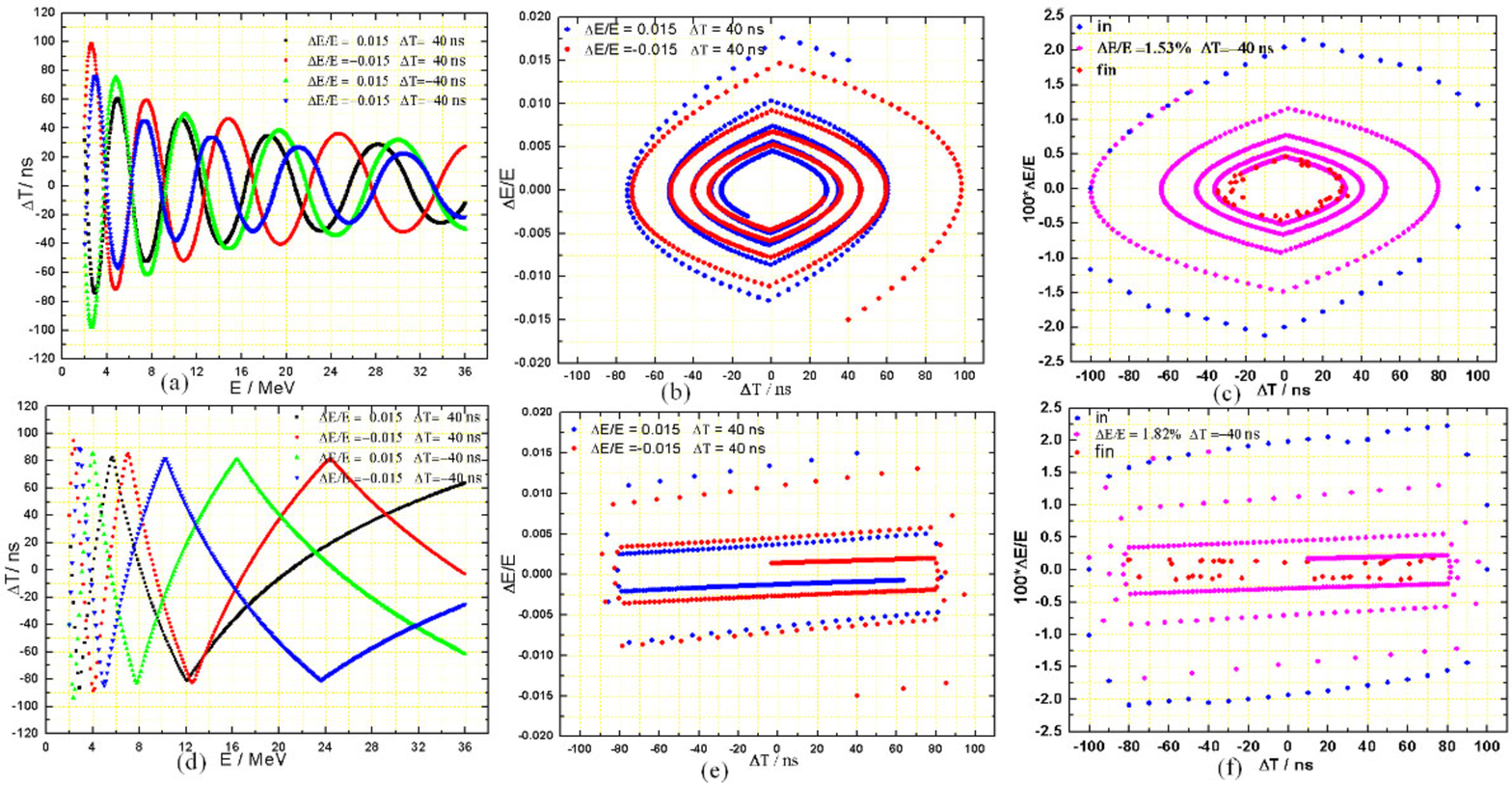}
\figcaption{\label{fig7} The evolution of accelerating time slip($\Delta$T) and energy deviation for some particles with different initial time (initial phase) and different initial energies on two constraint waveforms, (a)-(c) corresponding to double platform waveform and (d)-(f) corresponding to the head-tail restraint waveform. The accelerating time slip evolution with energy of four particles with initial time(initial phase) ¡À40 ns and initial energy spread ¡À1.5$\%$ are shown in (a) and (d) and the energy dispersion evolution is shown in (b) and (e). The particle longitudinal acceptance without space charge effect is given in (c) and (f).}
\end{center}
\begin{multicols}{2}
\vspace{-4mm}

In Fig.~\ref{fig7}, the results (a), (b) and (c) correspond to the simulation of double platform waveform and the results (d), (e) and (f) correspond to the simulation of head-tail restraint waveform. One assumes that some particles always in the accelerating interval [-100 ns, 100 ns] are never lost in the simulation. Without any particles loss, the relationship between the accelerating time (phase) and the maximum energy deviation on the injection time and extraction time is given in Fig.~\ref{fig7} (c) and Fig.~\ref{fig7} (f). The area surrounded by some blue dots can be approximate as particle longitudinal acceptance on the injection time. From the area surrounded by some red dots, one can see the particles maximum energy deviation and bunch length on the extraction time. The pink dots show an evolutionary process of energy deviation from the injection time to the extraction time for a particle.

The results show that the accelerating time slip($\Delta$T) and energy deviation of phase particles are all oscillating and the convergent processes toward the center time and ideal particles and also the convergence speed is becoming more and more slowly. Particles with energy deviation within ¡À1.5$\%$ and bunch length 200 ns are injected. When extracted corresponding to particle energy 36 MeV, particles after double platform waveform have energy deviation within ¡À0.5$\%$ and bunch length about 75 ns and particles after head-tail restraint waveform have energy deviation within ¡À0.25$\%$ and bunch length about 185 ns. Thus it can be seen that two constraint waveforms all provide the strong constraint on energy deviation and the double platform waveform provides stronger beam longitudinal focusing than head-tail restraint waveform. However, particle bunch excessively compressed in the longitudinal direction will result in the serious space charge effect and then affect the stability of accelerating voltage. Therefore, a very proper voltage difference of accelerating pulse waveform corresponding to the longitudinal focusing intensity in accelerating process is very important.

\section{Magnet design}
The magnetic field configuration design of combinatorial bending magnets for FFAG accelerator is one of the key problems. The radial magnet draws support from reverse bending magnet to balance the focusing effect of the horizontal and vertical directions, which is different from a spiral magnet with the aid of the edge effect providing a vertical focusing force. So the radial magnet has a simple magnet structure. The magnetic field configuration can be expressed as $B=B_{0}(r/r_{0})^{k}$, where $k$ is the field index shown in Table~\ref{tab1}. The field gradient is optimized by adjusting the magnet pole face shape and exciting current. H type bending magnet with a gap 62 mm is selected. The 3D model of a DFD super-period magnet is shown by OPERA-TOSCA in Fig.~\ref{fig8}.

\begin{center}
\includegraphics[angle=0,width=5.0cm]{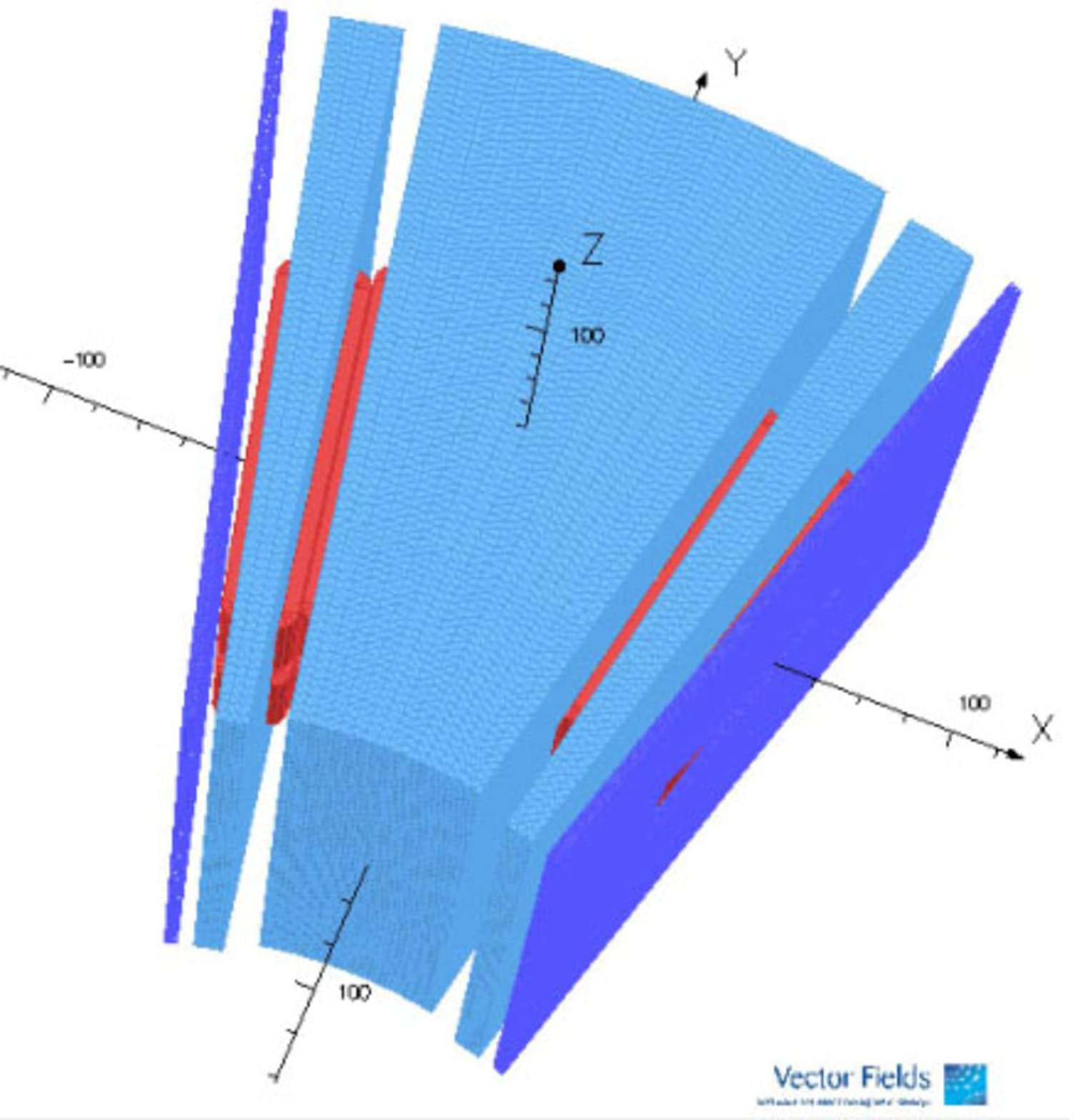}
\figcaption{\label{fig8} The magnet structure diagram of a super-period. The coil is shown by red and the shielding bodies are shown by purple in this figure.}
\end{center}

The field index and its relative change along the radial direction for focusing $F$ iron and defocusing $D$ iron on the middle plane are shown by Poisson Superfish in Fig.~\ref{fig9}. The magnetic field evolution of different arguments along the radial direction for focusing $F$ iron and defocusing $D$ iron is given in Fig.~\ref{fig10}. As shown in Fig.~\ref{fig10}, the magnetic field absolute value close to the magnet edge is less than near the magnet center area for each iron and also the more close to the edge, the more serious this phenomenon. Although magnet pole face near the edge is largely optimized to alleviate this phenomenon, the magnetic flux leakage and field interference still can't be ignored for the accelerating cavity and other equipments. So, shielding measures are necessary and reasonable.

\begin{center}
\includegraphics[angle=0,width=8.0cm]{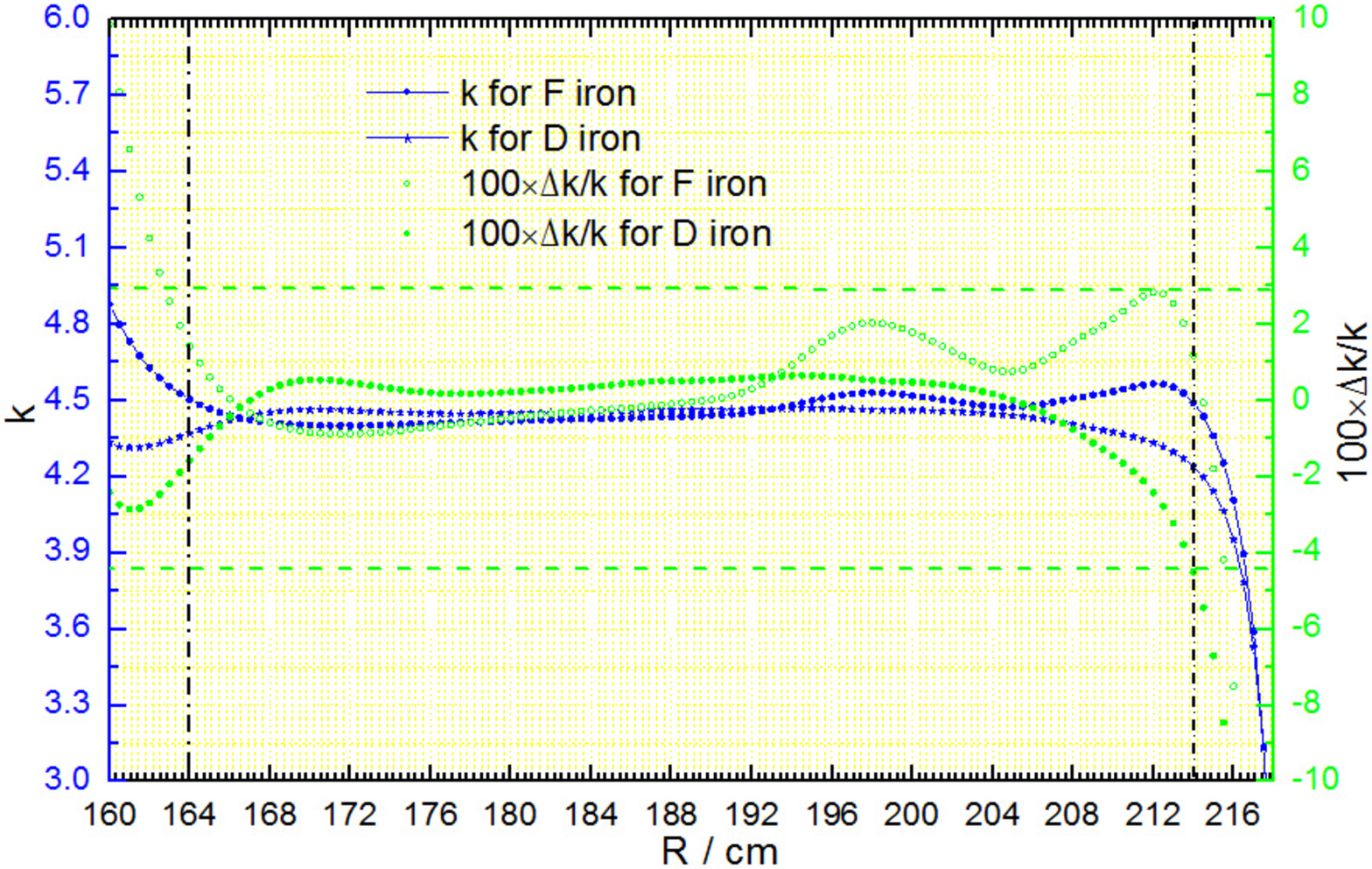}
\figcaption{\label{fig9} The field index and its relative change of the median plane in Poisson Superfish. The particle movement region is 164-214 cm, where the relative change of field index for two irons was optimized within 4.4$\%$.}
\end{center}

\begin{center}
\includegraphics[angle=0,width=7.5cm]{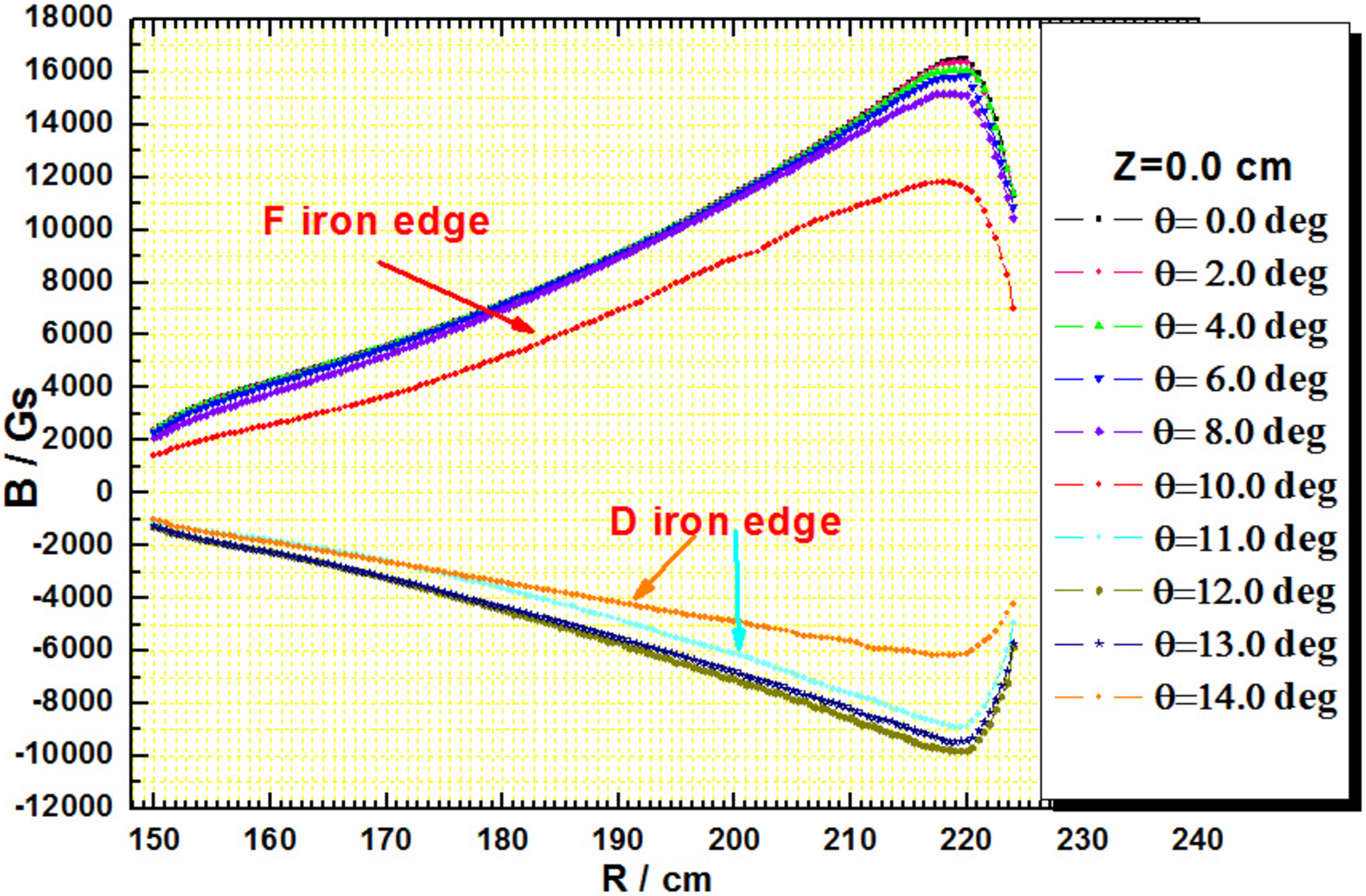}
\figcaption{\label{fig10} Magnetic field along the radial direction for different arguments in OPERA-3D. The half super-period is studied. Some lines marked by three arrows refer to the magnet edge.}
\end{center}

\section{The preliminary consideration for the beam injection and extraction scheme}
The scaling triplet DFD gives longer straight sections, which is beneficial in the assembly of the beam injection and extraction systems. For the accelerator scheme mentioned above, the average power of the beam with the maximum extracted energy 36 MeV and average beam intensity 10 mA is about 360 kW and the corresponding quantity of electric charge for the extraction repetitive frequency 500 Hz is 2 $\times$ $10^{-5}$ coulomb.

Considering the long-term steady and repeatable operation, 2.45 GHz microwave ion source with permanent magnet\cite{lab15} is adopted to produce 15 mA He$^{+}$ with a rms emittance 0.1 $\pi\cdot mm \cdot mrad$ approximately. The 30 mA He$^{+2}$ obtained by stripping an electron from 15 mA He$^{+}$ beam is accelerated by using high pressure platform or a small booster to 2 MeV and injected to helium ion FFAG accelerator with the help of septum and kicker magnets.

In order to improve the injected accomulating beam intensity, the multi-turn overlying pulse injection scheme with the number of pulses about 574 and injection time about 0.67 ms is adopted. For the helium ion FFAG accelerator, the time from the 2MeV beam injection to the maximum energy extraction is about 0.88 ms. The once extraction in single lap is considered because of the variable energy characteristic of FFAG accelerator.

\section{Summary}

The linear beam optics and combinatorial magnet design of helium ion FFAG accelerator are discussed in the paper. The radial scaling DFD scheme is adopted for the linear beam optics and the preliminary design is displayed. The magnet model with eight super-periods is given by OPERA-3D and also the magnetic field is further analysed. Still, more work is still needed to study the space charge effect and fringe effect and field disturbance caused by big magnet gap and aperture. The preliminary consideration for the beam injection and extraction scheme is also given in the fifth chapter.

The induction acceleration is applied to helium ion FFAG accelerator, instead of conventional radio frequency (RF) acceleration. The preliminary design value of some main parameters for induction cavity is listed through the theoretical analysis and also the cavity 3D model is presented by OPERA-3D. Two special constraint waveforms are proposed to solve the acceleration time slip($\Delta$T) phenomenon caused by accelerating voltage drop of flat top and the particle longitudinal motion in two waveforms is simulated. Some simulations show that without the limitation of space charge effect, double platform waveform and head-tail restraint waveform all can improve the constraint intensity on energy deviation and the former can provide stronger beam longitudinal focusing than the latter. So, the energy deviation constraint and longitudinal focusing of accelerated particles in helium ion FFAG accelerator are possible to simultaneously achieve by induction stacking technique of MOSFET solid-state modulator, which reduces or avoids the loss of accelerated particles. Moreover, it is a problem worth noticing in the future on how high intensity effects waveform.

\end{multicols}

\vspace{-1mm}
\centerline{\rule{80mm}{0.1pt}}
\vspace{2mm}

\begin{multicols}{2}

\end{multicols}

\clearpage

\end{document}